\documentclass[aps,twocolumn,showpacs]{revtex4}

\usepackage{hyperref,graphicx}

\begin{document}

\title{\bf  Accelerated dynamics with the dynamical
activation-relaxation technique}

\author{G.T. Barkema}
\email{barkema@phys.uu.nl}
\affiliation{
        Institute for Theoretical Physics, 
        Utrecht University, Leuvenlaan 4, 3584 CE Utrecht, 
        the Netherlands}

\author{Normand Mousseau}
\email{normand.mousseau@umontreal.ca}
%\homepage{http://www.esi.umontreal.ca/~mousseau}
\affiliation{
        D\'epartement de Physique and RQMP,
        Universit\'e de Montr\'eal,
        C.P. 6128, Succursale Centre-ville Montr\'eal,
        Qu\'ebec, Canada H3C 3J7.}

\date{\today}

\begin{abstract}
  The dynamics of many atomic systems is controlled by activated events
  taking place on a time scale which is long compared to that associated with
  thermal vibrations.  This often places problems of interest outside
  the range of standard simulation methods such as molecular dynamics.
  We present here an algorithm, the dynamical activation-relaxation
  technique (DART), which slows down thermal vibrations, while leaving
  untouched the activated processes which constitute the long-time
  dynamics.  As an example, we show that it is possible to accelerate
  considerably the dynamics of self-defects in a 1000-atom cell of {\it
  c}-Si over a wide range of temperatures.
\end{abstract}


\maketitle

Developing algorithms that stretch the time scale accessible to computer
simulations has been a major challenge in computational physics,
chemistry and biology. Standard techniques such as molecular dynamics
are strongly constrained by the presence of high-frequency modes in dense
materials and must therefore use an integration time step on the order of
a femtosecond, limiting the simulation length to microseconds at best.
Over the years, a number of accelerated algorithms have been proposed
for discrete systems where it is possible to enumerate, in advance,
all the possible barriers. The original ideas for these algorithms are
due to Bortz, Kalos and Lebowitz~\cite{bortz75}. They were first applied
to MBE growth in 1986 by Voter~\cite{voter86}. This method and many
variants are now common simulation techniques in surface science.

Algorithmic progress has been much slower for systems in which either
the number of pathways, or their activation energies, change all the
time.  Many methods have been proposed for sampling events efficiently
~\cite{barkema96,mousseau98,malek00,doye97,henkelman99,laio03,bolhuis00,thwart}
but only a few of these can provide a time scale: hyper-molecular
dynamics and related approches~\cite{voter97,miron03} as well
as temperature-assisted dynamics, introduced by Sorensen and
Voter~\cite{sorensen00}, are limited to systems with a relatively narrow
distribution of barriers and cannot be used easily at high temperatures
or on generic problems. Other methods, such as a kinetic Monte Carlo
scheme with on-the-fly calculation of the barriers~\cite{henkelman01}
generate impressive acceleration.  However, their application is limited
to relatively simple systems, and entropic effects are not fully included,
hence detailed balance cannot be ensured.

In this paper, we present an algorithm, the dynamical
activation-relaxation technique (DART), with a self-correcting
accelerating factor that can reach a few orders of magnitude at
physically relevant temperatures while overcoming some of the
limitations of these accelerated methods.  It is based on the
thermodynamically-weighted activation-relaxation technique (THWART)
which we introduced recently~\cite{thwart}, and combines molecular
dynamics with Monte Carlo methods to reach a dynamically correct
acceleration of the slow processes in complex materials. In order to
assess the efficiency of DART, we apply the method to the diffusion of
vacancies and interstitials in {\it c}-Si.

\begin{figure}
\includegraphics[width=7cm]{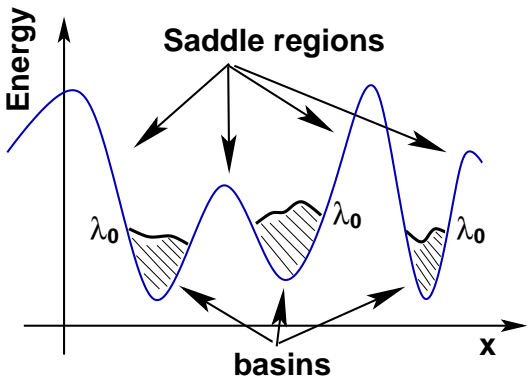}
\vskip 6pt
\caption{The energy landscape of systems with a long-time dynamics 
  determined by activated events.}
\label{fig:landscape}
\end{figure}

The energy landscape of systems with dynamics controlled by rare
events can be divided into two types of subregions, as shown in
Fig.~\ref{fig:landscape}: the basins, closed regions of the energy
landscape, in which the system is confined for extended periods and
which can contain many local minima, and the ``activated part of phase
space'', sampled only when a rare fluctuation pushes the system from one
basin to another.  Following THWART, we delineate the basins based on
the value of the lowest local curvature of the energy landscape (i.e.,
the lowest eigenvalue of the hessian matrix); any configuration with a
lowest curvature below the threshold is considered to be in the saddle
region.  The exact threshold value depends on the system
studied as well as on the simulation temperature.

The simulation starts with standard molecular dynamics in a
microcanonical ensemble. After equilibration, the MD run is pursued
but with regular evaluation of the lowest local curvature $\lambda_0$.
As long as $\lambda_0$ remains above a threshold value $\lambda_t$,
the system is considered to reside in the original basin. As
soon as this threshold is reached, however, the activation phase of
THWART is launched: the MD simulation is stopped, the current
velocities stored and the system is moved from one basin to another.
The step is designed to bring the configuration from the edge of a
basin to that of a neighboring one, in a fully reversible manner and
at constant potential energy in order to respect detailed balance; it
is identical to that used in THWART. Each activation move consists of
a sequence of small steps with size $\Delta x$, defined as follows:
\begin{equation}
\vec{x}_{i+1} = \vec{x}_i + \frac{\Delta x}{2} \left( \vec{h}_{i+1} +
\vec{h}_{i}\right) + c \left( \vec{F}^{\perp}_{i+1} + 
\vec{F}^{\perp}_i \right),
\label{eq:thwart}
\end{equation}
where $\vec{h}$ is the direction of lowest curvature and $\vec{F}^{\perp}$
is the force perpendicular to this vector; $c$ is a scalar tuned
at each step to keep the total configurational energy constant.
The orientation of $\vec{h}_0$ is chosen away from the basin, while
successive orientations are chosen such that $\vec{h}_{i+1} \cdot
\vec{h}_i >0$. This step is repeated until the lowest curvature of
the energy landscape reaches again the threshold value $\lambda_t$.
Note that each individual step, and hence the whole trajectory, is fully
reversible.  Since each activation move connects two points in phase space
with the same potential energy and since the Jacobian of transformation
is equal to one~\cite{thwart}, detailed balance is fully respected.
Within THWART, activated moves are always accepted; the MD is therefore
continued from the end of the activation path using the stored velocities,
ready for a new activation.

THWART ensures a proper thermodynamical sampling of the phase space;
the THWART trajectory, however, does not follow the real dynamics.  In
particular, it crosses highly activated pathways as easily as those
with a low activation barrier.  In order to recover the actual
dynamics, it is necessary to first determine the height of the
barriers crossed.  By construction, activation pathways have constant
energy and the deformation energy stored in the degrees of freedom not
directly involved in the activation process is used as a bath to
enforce this constraint. It is nevertheless straightforward to 
to determine the energy associated with the diffusion path: the total
force is split into a component parallel to the direction of lowest
curvature ($\vec{F}^{\parallel}$) and a component perpendicular to
this direction ($\vec{F}^{\perp}$), at each step along the activation
trajectory; using the first component, the
change in total energy due to the parallel displacement is written
\begin{equation}
\Delta E^{\parallel} \equiv - \int \vec{F}^{\parallel} \cdot d\vec{R}.
\end{equation}
The activation energy barrier (from the edge of the basin) is then
defined as the change in energy from the edge of the basin to
the maximum energy change projected along this path.

The probability that the kinetic energy at the beginning of the activation
trajectory suffices to bring the configuration over the barrier through
this highest point is given by $\exp(-\beta \max (\Delta E^{\parallel}))$,
with inverse temperature $\beta=1/(k_bT)$. 
Assuming around each first-order saddle point
a quadratic behavior of the energy in the transition plane, the energy at the
nearest saddle point will on average be $\frac{1}{2} k_bT$ lower. We therefore
take for the activation barrier, faced from the basin boundary:
\begin{equation}
\Delta E_{\rm act} = \max (\Delta E^{\parallel}) - \frac{1}{2} k_bT.
\label{eq:act}
\end{equation}
To retrieve the dynamics in a statistically correct manner, we should
accept the activation move with probability $\exp(-\beta \Delta E_{\rm act})$,
and otherwise continue in the original basin.  Typically in a system
with activated dynamics, these acceptance probabilities are rather small,
and the system will bounce back and forth in the basin many times before
eventually escaping from it; hence the slow dynamics. However, the speed
of the simulation can be enhanced by a (constant) nominal boost factor $X_b$
reducing the number of such bounces. The acceptance
probability for an activation pathway then becomes
\begin{equation}
P_{\rm cross} = \min \left[ 1, X_b \exp \left( -\beta 
\Delta E_{\rm act} \right) \right].
\label{eq:cross}
\end{equation}

If, on top of boosting the acceptance probabilities by a factor of $X_b$,
the time scale is stretched by the same factor, the long-time dynamics is
untouched, provided it is indeed determined by activated processes with
barriers exceeding $\ln(X_b) k_bT$, while the suppression of the in-basin
dynamics (by a factor $X_b$) as well as the suppression of less activated
processes (by a smaller factor) does not affect the long-time dynamics.
Once barriers below $\ln(X_b) k_bT$ are encountered, inevitably some
distortion of the dynamics occurs. To alleviate the distortion somewhat,
if we encounter such a low barrier, we stretch the time scale since
the previous event by an on-the-fly corrected boost factor
\begin{equation}
X_{\rm eff}=P_{\rm cross} \cdot \exp(\beta \Delta E_{\rm act}),
\end{equation}
rather than the nominal boost factor $X_b$; this recovers correct
dynamics for systems in which the activation energy is constant, even
if the chosen nominal boost factor is too large.

A DART simulation proceeds in the following sequence: (1) at time $t=0$,
a microcanonical molecular dynamics simulation is launched and the
value of the lowest local curvature $\lambda_0$ is monitored at regular
intervals (typically, every 50 steps); (2) when $\lambda_0$ reaches the
threshold value $\lambda_t$, the molecular dynamics is stopped and the
velocities are saved; (3) following Eq.~(\ref{eq:thwart}) iteratively,
the configuration is brought into a new basin and the activation energy
$\Delta E_{\rm act}$ is computed along the activated pathway; (4) the
event is accepted with probability $P_{\rm cross}$; (5) if the event is
accepted, the time is incremented by $\Delta t=t_{\rm MD} \cdot X_{\rm
eff}$, in which $t_{\rm MD}$ is the time spent doing molecular dynamics
since the previous accepted event, and the molecular dynamics simulation
is continued starting at the new edge with the same velocities. If it
is rejected, the molecular dynamics simulation is simply continued from
the initial edge.

To demonstrate the efficiency of DART, we consider the diffusion of
vacancies and interstitials in {\it c}-Si, described by the
Stillinger-Weber potential. Both types of defects have been well
characterized previously~\cite{maroudas93,nastar96}. Vacancy diffusion
is associated with a single activation barrier of 0.43
eV~\cite{maroudas93}.  An interstitial can take four different stable
topologies: tetrahedral, hexagonal, bond-centered and split or
dumbbell. It diffuses through many mechanisms with activation barriers
between 0.65 and 1.62 eV~\cite{nastar96}. Therefore, these two defects
provide us with various levels of complexity to test DART.

\begin{figure}
\includegraphics[width=7cm]{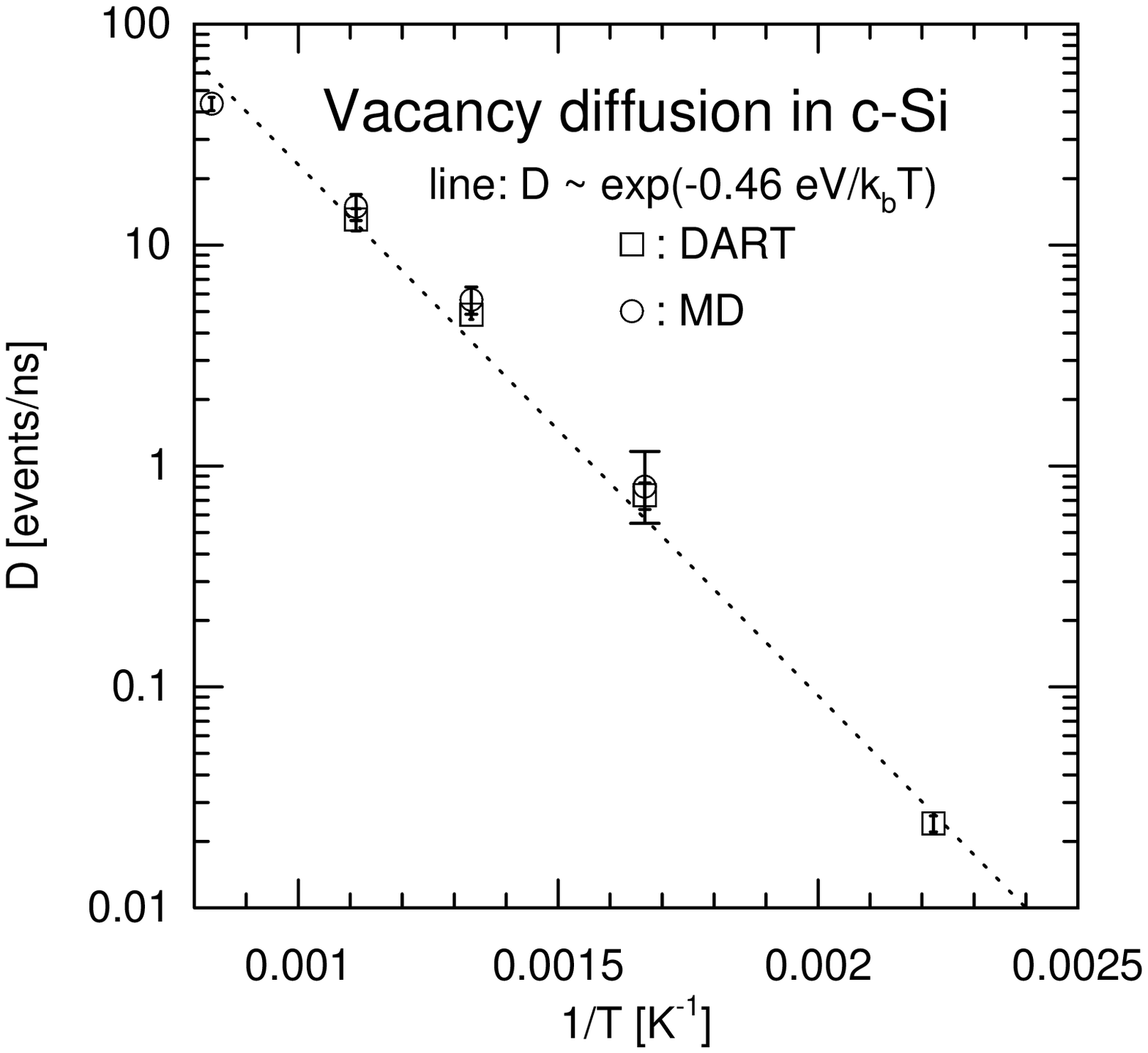}\\
\includegraphics[width=7cm]{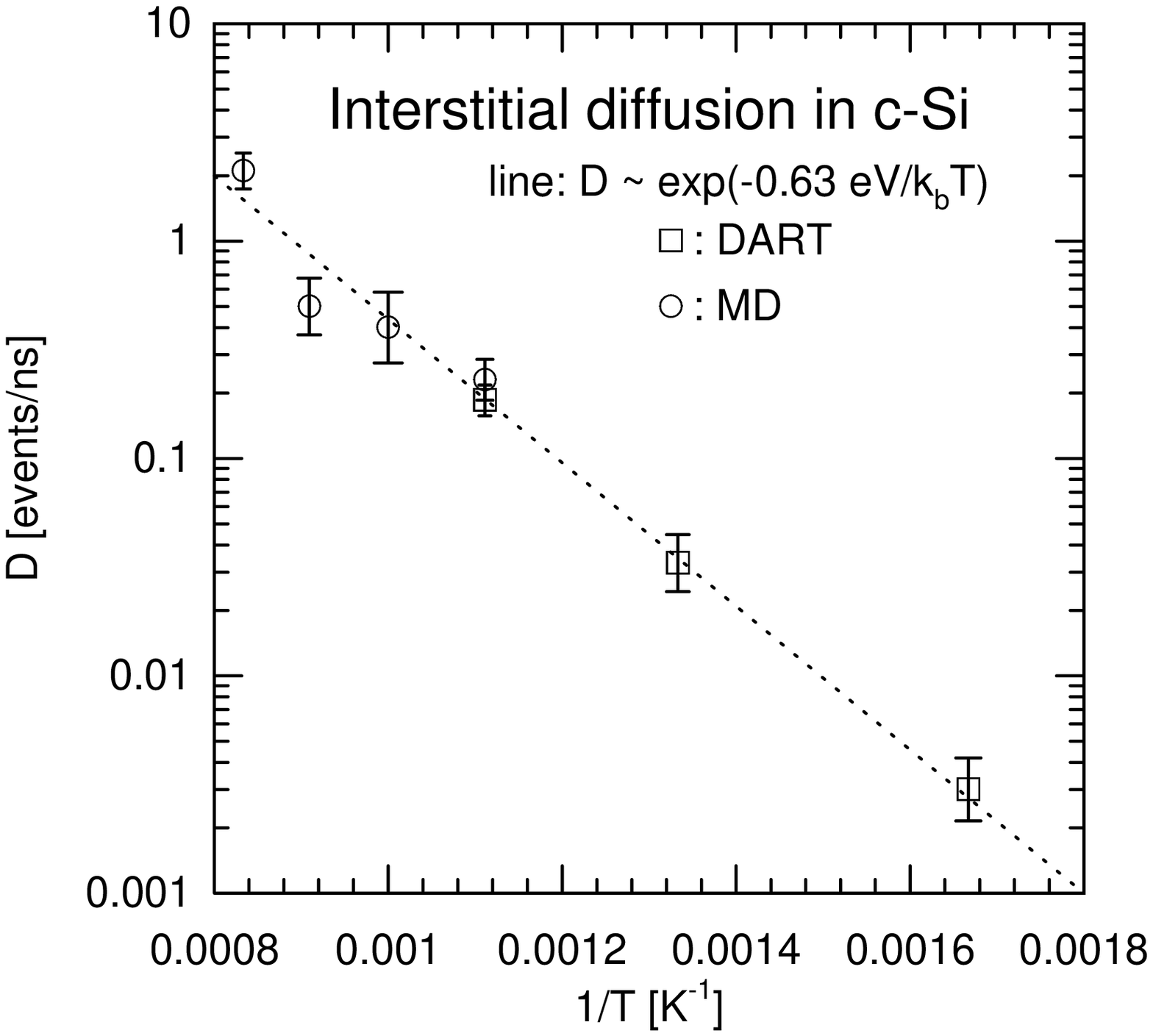}
\caption{ Diffusion rate as a function of inverse temperature for vacancy
(top) and the interstitial (bottom) diffusion in Si. The open circles
represent MD results and the squares DART results.  Error bars are
obtained from the square root of the total number of events.  The nominal
and averaged effective boost factors for these simulations are given in
Table ~\ref{tab:boost}. }
\label{fig:boost}
\end{figure}

Figure ~\ref{fig:boost} shows an Arrhenius plot of the diffusion rates
obtained by MD and by DART with various boost factors.  We characterize
the diffusion rates by the hopping rates of the defects, which gives
better statistics than the mean squared displacements per unit of time.
As can be seen, an excellent agreement between the two methods is achieved
for both the vacancy, characterized by a single energy barrier, and for
the interstitial, which shows a more complex behavior.

As mentioned above, DART adjusts automatically for barriers lower than
$\ln(X_b) k_bT$. Table ~\ref{tab:boost} gives the nominal and
effective boost factors for the various DART simulations plotted in
Fig.\ ~\ref{fig:boost}. As expected, the gain in efficiency increases
rapidly with decreasing temperature. From a factor of 1 at about 1100K
(not shown), the effective boost for interstitial diffusion reaches 10
at 900 K, 74 at 750 K and almost 150 at 600 K, an experimentally
relevant temperature.   The extra computational effort in DART is
largely due to the computation of the lowest curvature, both at
regular intervals and during the THWART events. Averaged over the
whole simulation, with an evaluation at every 50 MD steps, a DART time
step costs slightly less than two MD time steps.

\begin{table}
\caption{Details of the various DART simulations}
\begin{tabular}{lcccc}
Defect & Temperature & $\lambda_0$ & 
\multicolumn{2}{c}{Boost} \\
&(K)& (eV/ \AA$^2$) &Nominal & Effective \\ \hline \hline
Vacancy
        & 900  & -5& 6 & 3.6\\
        & 900  & -5& 30 & 6.7\\
        & 600 & -5& 6 & 5.4\\
        & 600 & -5 & 30 & 19 \\
        & 450 & -5 & 600 & 271 \\
\hline
Interstitial
             & 900  & -7 &  60 & 10\\
             & 750 &  -7& 600 & 74\\
             & 600 &  -7& 1200 & 148\\
\hline
\end{tabular}
\label{tab:boost}
\end{table}

In conclusion, we have presented here an accelerated molecular
dynamical method -- the dynamical activation-relaxation method (DART).
This algorithm provides a tunable acceleration parameter that can be
adjusted to suit the specific problems studied. In addition to
providing a significant acceleration over MD, DART has numerous
advantages: (1) the algorithm is not very sensitive to the various
parameters - it can automatically correct for boost factors an order
of magnitude or more too large; (2) it computes relaxation
trajectories and activation barriers {\em on the fly}, leading to a
very low overhead (on average, a DART time step is less than twice the
cost of an MD time step); (3) DART is not slowed down by the local
rearrangements which take place in the basin (i.e, below threshold) ---
contrary to other accelerated methods, it is therefore possible to use
DART to accelerate the dynamics of more complex systems such as
glasses and proteins; (4) since the events are easily labeled, it is
possible to use various tricks to avoid repeating the same event over
and over again -- this can lead to a large increase in efficiency;
finally, (5) the limits of DART are well behaved: a zero-boost reduces
DART to standard MD, while an infinite boost factor recovers THWART
and still ensures proper thermodynamical sampling.

Tests on a vacancy and an interstitial in {\it c}-Si have shown that
this method remains accurate dynamically with an increased
computational efficiency of two or more orders of magnitude, at
temperatures as high as 450 K for the vacancies or 600 K for
interstitial diffusion. As mentioned above, however, this method can
also be applied to much more complex situations where current
accelerated algorithms fail.

{\it Acknowledgements.} NM is supported in part by FQRNT and NSERC.
We are grateful to the RQCHP for generous allocation of computer
resources. NM is a Cottrell Scholar of the Research Corporation.

\end{document}